\documentstyle[12pt,aaspp4]{article}

\begin{document}

\title{Star Formation in Las Campanas  Compact Groups}

\author{Sahar S. Allam\altaffilmark{1}}
\affil{National Research Institute for Astronomy \& Geophysics, 
Helwan, Cairo, Egypt; shr@frcu.eun.eg,
sallam@fnal.gov}
\altaffiltext{1}{Visiting Scientist, Fermi National Accelerator 
Laboratory} 

\author{Douglas L. Tucker}
\affil{Fermi National Accelerator Laboratory, MS 127, P.O. Box 500, 
Batavia, IL 60510, USA;
dtucker@fnal.gov}

\author{Huan Lin\altaffilmark{2}}
\affil{Steward Observatory, University of Arizona, 933 N. Cherry Ave. 
Tucson, AZ 85721, USA; hlin@as.arizona.edu}
\altaffiltext{2}{Hubble Fellow}

\and

\author{Yasuhiro Hashimoto\altaffilmark{3}}
\affil{Carnegie Observatories, 813 Santa Barbara Street, Pasadena, 
CA 91101, USA; hashimot@vorpal.ociw.edu}
\altaffiltext{3}{Also:  Dept.\ of Astronomy, Yale University, New Haven, CT
06520-8101; hashimot@astro.yale.edu}

\begin{abstract}

Compact groups (CGs) of galaxies offer an exceptional laboratory for
the study of dense galaxian environments --- where interactions,
tidally induced activity, and mergers are expected to be at their
highest rate of occurrence.  Here, we present first results from a new
catalogue of compact groups, one based upon the Las Campanas Redshift
Survey (LCRS). Using the equivalent width of [\ion{O}{2}]\,
$\lambda$\,3727, we have studied the star formation activity in LCRS
CGs: we find strong evidence of {\it depressed star formation} in CGs
relative to that in loose groups or the field. Although much of this
effect can be ascribed to morphological mix (CGs contain a high
fraction of early-type galaxies), there is some evidence that the star
formation rate in late-type galaxies is particularly deficient --- 
perhaps only one-half to one-third that of field spirals. We conclude
that gas stripping mechanisms may play a role in CG environments.
\end{abstract}


\section{Introduction}

Perhaps over half of all galaxies lie within groups containing 3--20
members (\cite{Tully87}); yet, due to the difficulty of discerning
them from the field, groups of galaxies are, as a whole, not as well
studied as larger galaxy systems.  Compact groups (CGs), however,
defined by their small number of members ($< 10$), their compactness
(typical intra-group separations of a galaxy diameter or less), and
their relative isolation (intra-group separations $\ll$ group-field
separations) are more readily identifiable.

Recently, Tucker et al.\ (1999) produced a catalogue of loose groups
(LGs) from the Las Campanas Redshift Survey (LCRS;
\cite{Shectmanetal96}), using an adaptive friends-of-friends algorithm
(\cite{Ramellaetal89}). Intrigued by the work of \cite{Bartonetal96},
who created a CG catalogue from the Center for Astrophysics (CfA)
Redshift Survey and found that most of their CGs were embedded in
dense environments, we produced a similar catalogue from the much
deeper LCRS (\cite{AllamTucker98}, \cite{Tuckeretal99}).  For
extracting group catalogues, redshift surveys have an advantage over
sky surveys since redshift adds a third dimension of constraint: group
catalogues based upon redshift surveys tend to have far fewer chance
alignments than do those based upon sky surveys (e.g.,
\cite{Hickson82}, 1993; HCG). We apply a standard friends-of-friends
algorithm to extract a sample of CGs systems in the LCRS.  Our
definition for these CGs is as follows:
\begin{itemize}
\vspace{-.15cm} 
\item $ \geq$ 3 galaxies, 
\vspace{-.25cm} 
\item compact  (projected nearest-neighbor inter-galaxy separations of $D_{L} 
\le$ 50$h^{-1}$kpc, or $\sim$ 1 galaxy diameter), and
\vspace{-.25cm}
\item isolated in redshift (nearest-neighbor inter-galaxy velocity
differences $V_{L} \leq$ 1000~km~s$^{-1}$).
\vspace{-.15cm} 
\end{itemize} 

The LCRS, optimized for efficient observing with a fiber-fed
multi-object spectrograph, has a 55~arcsec fiber separation limit.
This has prevented the observation of spectra for all galaxies which
were members of close pairs; so, many galaxies in CG environments are
missing from the LCRS redshift catalogue.  We have partially
circumvented this problem by assigning each of the $\sim$1,000
``missing'' LCRS galaxies the redshift of its nearest neighbor and
convolving it with a gaussian of $\sigma$=200~km~s$^{-1}$, a value
which is similar to the typical median velocity disperion of HCGs
(Hickson 1982) and of LCRS LGs (Tucker et al.\ 1999); hence, on the
small angular scales necessary for compact group selection, the LCRS
falls somewhere between a 2D sky survey and a fully 3D redshift
survey.  The resulting catalogue contains 76 CGs having 3 or more
members, and evidence for interactions in many of these CGs (in the
form of tidal tails, bridges, etc.; see \cite{AllamTucker98},
\cite{Allametal99}) confirms that they are indeed, for the most part,
physical systems.  All the CGs contain at least one redshift; 23
contain 2 or more. (Unfortunately, only one LCRS CG has redshifts for
all its members.)  The innate physical properties of LCRS CGs --- such
as typical group richnesses and densities --- are similar to those of
the Barton et al.\ catalogue, which in turn are similar to those of
the HCG catalogue, especially for CGs with 4 or more members.  The
median redshift for LCRS CGs, however, is $\sim$0.08, more than twice
that of either of the other two CG catalogues.  As with the HCG and
Barton et al.\ samples, LCRS CGs represent some of the densest
concentrations of galaxies known and thus provide ideal laboratories
for studying the effect of strong interaction on the morphology and
stellar content of galaxies.  Details of the general properties of
these CGs and of how they were extracted from the LCRS will be
discussed in Allam et al.\ (1999); here, we will focus on the star
formation properties in LCRS CG environments.

It is well known that direct interactions between galaxies tend to
increase their star formation rate (SFR) (\cite{LarsonTinsley78};
\cite{Bushouse87}; \cite{Kennicuttetal87}).  LCRS CGs represent an
environment where interactions, tidally triggered activity, and galaxy
mergers are expected to be at their highest rate of occurrence.
Therefore, if no other factors dominate, we may expect a global
enhancement in the SFR of LCRS CG galaxies.  In order to test this
hypothesis, we will use the equivalent width (EW) of the
[\ion{O}{2}]\,$\lambda$\,3727 emission line (\cite{Collessetal90},
\cite{Kennicutt92}) as a star formation indicator.

The paper is organized as follows: \S~2 describes the sample
under investigation, \S~3 discusses the sample's spectroscopic
properties, and \S~4 relates the sample's morphological
features; finally, in \S~5, we summarize our main conclusions.

\section{The Samples}

As a first step towards the clarification of the effect of high
density environments on enhancing the SFR in galaxies, it is necessary
to characterize the SFR of galaxies in more isolated environments. For
that reason, a sample of 253 CG galaxies, a sample of 7621 LG
galaxies, and a sample of 13452 field galaxies have been selected from
the LCRS.  Particular care was taken in order to obtain a loose group
sample in which no galaxies from CGs were included.  Further, galaxies
from both LGs and CGs were excluded from the field sample.  Our goal
is to study environmental factors affecting the SFR of galaxies by
taking advantage of the very large and homogeneous data set available
from the LCRS.

Before we move on, however, a concern must be addressed: could the
fiber separation effect --- the fact that, in high-density regions,
the fraction of LCRS galaxies with spectra is lower than that in
low-density regions --- bias our analysis?  To first order, this
concern is unimportant, since we are comparing the fraction of
starbursts (see \S~3) against the total sample of galaxies with
spectra --- not against the total sample of galaxies both with and
without spectra. Furthermore, the galaxies removed due to the fiber
size were removed blindly --- i.e.\ with no regard to their star
formation properties or morphological type.  On the other hand,
uncertainties in group membership due to the fiber separation effect
can obscure the boundary between low- and high-density regimes,
possibly diluting the differences in the observed properties of these
environments.  In other words, any environmental effects we detect
would likely be even stronger in an uncontaminated sample.

\section{Distribution of [\ion{O}{2}]  Equivalent Widths}

\begin{figure}[htbp]
\plotfiddle{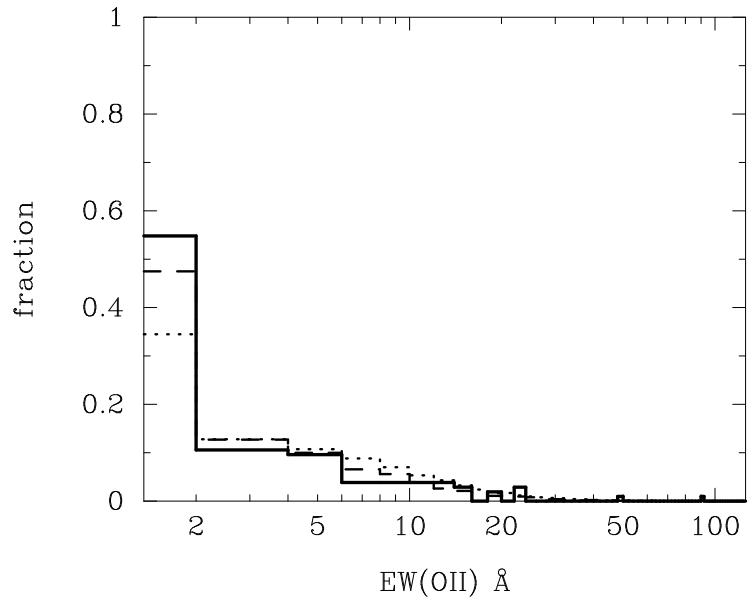}{1.50cm}{-90}{50}{50}{-200}{100}
\vspace{7.75cm}
\caption{The distribution of the equivalent widths (EW) of
[\ion{O}{2}] $\lambda$3727 of LCRS galaxies in compact groups (full
line), loose groups (dashed line), and the field (dotted line).  A
formal $\chi^{2}$ test indicates that the distribution for compact
groups differs from that for loose groups at the 99.99965\% confidence
level, and from that for field galaxies at the 99.99951\% confidence
level.  \label{fig.EW.fraction}}
\end{figure}

Several works have used EW(\ion{O}{2}) \,$\lambda$3727 as a star
formation index for distant galaxies (\cite{Collessetal90},
\cite{Kennicutt92}).  We have used automatically measured rest-frame
LCRS EW(\ion{O}{2})'s, which have a mean error of 2.2~\AA \,
[\cite{Hashimotoetal98}].  \placefigure{fig.EW.fraction}
Figure~\ref{fig.EW.fraction} shows the distribution of the
EW(\ion{O}{2}) of LCRS galaxies in CGs, in LGs, and in the field.  A
formal $\chi^{2}$ test indicates that the distribution for CGs differs
from that for LGs at the 99.99965\% confidence level, and from that
for field galaxies at the 99.99951\% confidence level.  (These very
high formal confidence levels are due partly to the large samples
involved and partly to the large differences among these samples for
the smallest bin.)

Following \cite{Hashimotoetal98}, we classify the emission line
strength as follows: NEM (no emission), for which EW$<$5$\AA$; WEM
(weak emission), for which 5$\AA$ $\leq$EW$<$20$\AA$; and SEM (strong
emission), for which EW$\geq$20$\AA$. The WEM class contains mostly
normal galaxies, where star formation is governed by internal factors
such as gas content and disk kinematics. The SEM class contains mainly
starburst galaxies, where star formation is due to interaction.  Table
\ref{tab:EW} represents the frequency of EW(\ion{O}{2}) for galaxies
in different environments. The variations in the frequency of the SEM
class may reflect environmental variations in galaxy-galaxy
interaction rates.  \placetable{tab:EW}

Note that the fraction of LG galaxies showing a normal (WEM) SFR is
only three-quarters that for the field galaxies, and the fraction of LG
galaxies showing starburst (SEM) activity is only two-thirds that in
the field. For CG galaxies, the ratios are more severe: the fraction
of CG galaxies with normal SFR is only two-thirds that for the field
galaxies, and the fraction of CG galaxies which are star-bursting is
only half that of the field, indicating that the SFR in high density
environments is generally weaker than in the field.

\section{The Concentration Index $C$ of LCRS galaxies}

Although the SFR in high density environments is, on average,
depressed relative to that than in the field, much of this effect
might be due merely to differences in average morphological mix.
After all, spirals, which are more prevalent in the field, tend to
have higher average SFRs than do ellipticals.  To test this
possibility, we have made use of \cite{Hashimotoetal98}'s measurement
of the concentration index, $C$, for LCRS galaxies as a measure of the
morphological types of the galaxies in our sample.  The $C$ index
represents the intensity-weighted second moment of a galaxy; it
compares the flux between specified inner and outer isophotes of a
galaxy to indicate the degree of light concentration. As such, the $C$
index is related to the Hubble type (\cite{Abrahametal94}), where
late/irregular type galaxies have smaller $C$ values.  The total number
of galaxies in our sample with a measured $C$ index is 12901. The mean
and median $C$ index is given for each of the different galaxy
environments in Table \ref{tab.C}.

\begin{figure}[h]
\plotfiddle{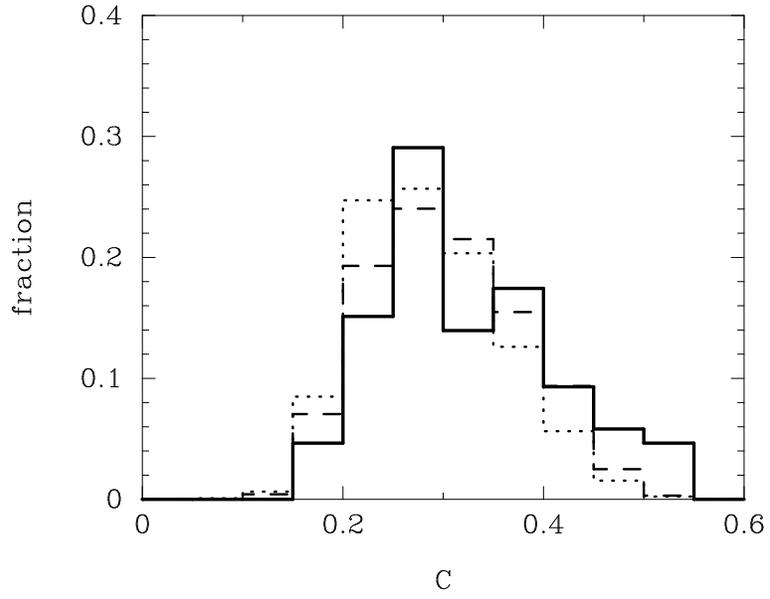}{1.50cm}{-90}{50}{50}{-190}{115}
\vspace{5.5cm}  
\caption{Distribution of the concentration index $C$ of LCRS
compact group galaxies (full line), loose group galaxies (dashed
line), and field galaxies (dotted line).  \label{fig.C}}
\end{figure}

\begin{figure}[htbp]
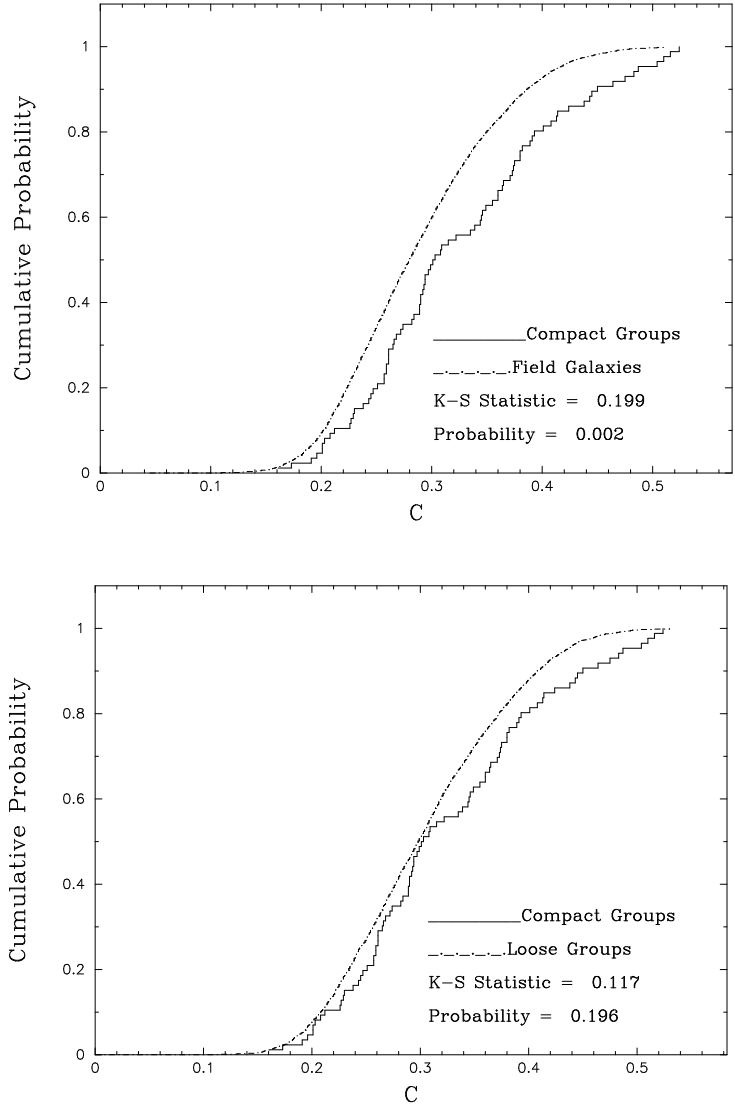

\plotfiddle{f03_top.ps}{1.50cm}{-90}{45}{45}{-180}{110}
\plotfiddle{f03_bot.ps}{1.50cm}{-90}{45}{45}{-180}{-50}
\vspace{12.5cm}  
\caption{The cumulative distribution of the concentration index
$C$ of the LCRS compact group galaxies vs.\ field galaxies (top), and
compact group galaxies vs.\ the loose group galaxies (bottom). The $C$
distribution of LCRS compact group galaxies is skewed toward early
type (large $C$).  \label{fig.ks.C}}
\end{figure}

\placefigure{fig.C}
\placetable{tab.C}
\placefigure{fig.ks.C}

\begin{figure}[htbp]
\plotfiddle{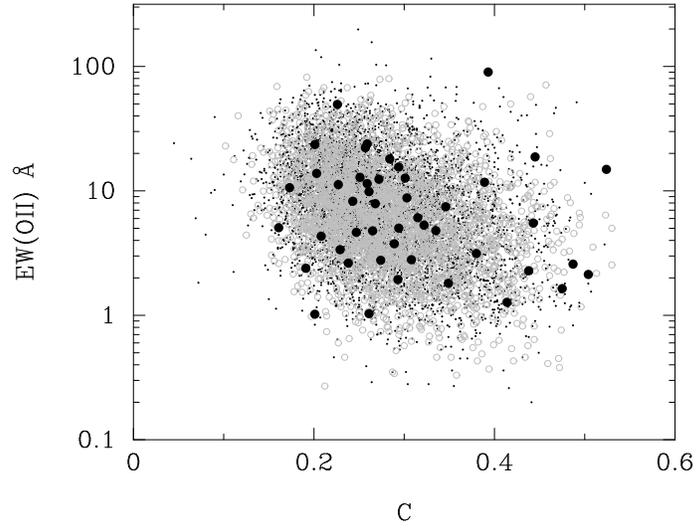}{1.50cm}{-90}{45}{45}{-180}{90}
\vspace{5.85cm}  
\caption{EW(\ion{O}{2})\,$\lambda$3727 vs.\ $C$ index for LCRS
galaxies in compact groups (filled circles), in loose groups (unfilled
circles), and in the field (points). \label{fig.EW.C}}
\end{figure}

The $C$ distribution of CGs galaxies is shown in Figs.~\ref{fig.C} \&
\ref{fig.ks.C}.  A KS test indicates that the CG galaxies are drawn
from the same morphological parent population as the LG galaxies at a
probability of 20\%; the probability that CG and the field galaxies
have the same morphological mix is only 0.2\%.  Clearly, the
distribution of CG galaxies is skewed toward early types (large $C$'s).
\placefigure{fig.EW.C} 

\begin{figure}[htbp]
\plotfiddle{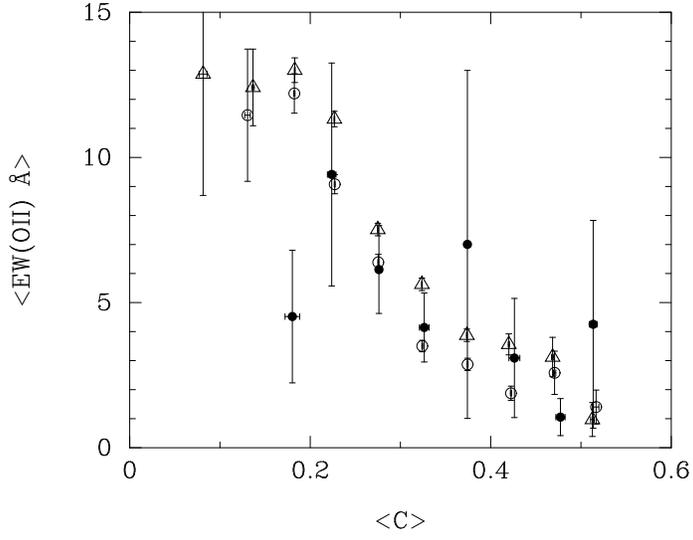}{1.50cm}{-90}{45}{45}{-250}{90}
\vspace{5.85cm} 
\caption{The relation between the mean concentration index,
$<$$C$$>$, and the mean EW(\ion{O}{2}), $<$EW(\ion{O}{2})$>$, for
compact group galaxies (filled circles), for loose group galaxies
(open circles), and for field galaxies (open triangles).
\label{fig.avC}}
\end{figure}

\placefigure{fig.avC}

In Fig.~\ref{fig.EW.C}, the distribution of EW(\ion{O}{2}) vs.\ $C$
index is shown for LCRS galaxies in the different environments.  The
relation between the mean $C$ index, $<$$C$$>$, and the mean
EW(\ion{O}{2}), $<$EW(\ion{O}{2})$>$, is presented in
Fig.~\ref{fig.avC}.  Note that $<$EW(\ion{O}{2})$>$ increases smoothly
with decreasing $<$$C$$>$ for LG and field galaxies, parallelling the
relation between Hubble type and EW(\ion{O}{2}) (\cite{Kennicutt92}).
Although much noisier, the same relation holds basically true for CG
galaxies, too.  We must note, however, that the latest-type (the
smallest $C$ bin) CG galaxies show a significant deficit of star
formation --- perhaps only one-half to one-third that of field
galaxies of this morphology.  Therefore, it appears that not all the
differences between the average star formation properties of CGs, LGs,
and the field are due merely to morphological mix.  Some appear to be
due to the dampening of star formation within late-type CG galaxies.

\section{Conclusion}

The star formation histories of galaxies in CGs can provide insight
into the environmental factors that influence the evolution of
galaxies.  One approach is to examine the spectra of galaxies for
evidence of ongoing star formation or of a young stellar
population. We can then compare the fraction of compact group galaxies
with recent star formation with the fraction from loose groups and the
field.

We have done this by making use of a new catalogue of CGs, based upon
the LCRS, which contains 253 galaxies in 76 CGs.  To clarify whether
interaction produces enhanced star formation in LCRS CGs, they have
been compared to carefully selected samples of LCRS LG and field
galaxies.  In all, a sample of 21326 LCRS galaxies in the three
different environments was employed.

We compared the SFR based on the strength of the emission line
EW(\ion{O}{2}) for LCRS CGs, LGs, and field galaxies: we found that the
fraction of starbursts for CG members is roughly half that for the
field, whereas for LG galaxies it is roughly two-thirds that for the
field.  Also we found that a normal galaxy SFR occurs for LCRS CG
galaxies at roughly two-thirds the rate for the field, whereas for LG
galaxies this rate is three-fourths that for the field. This means
that, on average, the star formation in high density environments is
depressed with respect to the field.
 
Much of this effect can be attributed to the different morphological
mixes associated with low and high density environments: when we
compared the distribution of the concentration index $C$ of galaxies
in CGs, in LGs, and in the field, we found the distribution of CGs
galaxies to be definitely skewed towards early morphological types
(large $C$ index), which generally tend to have relatively low SFRs.
Nonetheless, when we then compared the SFR vs.\ the $C$ index for CG,
for LG, and for field galaxies, we found that the SFR for CGs appears
to be deficient for very late morphological types (small $C$ index)
--- in fact, the SFR for these late-type CG galaxies is only one-half
to one-third the SFR for field spirals.

It is clear from these findings that CG environments tend to depress
star formation, partly due to a relative overabundance of early-type
galaxies and partly due to some mechanism that dampens star formation
within late-type CG spirals.  Note that results from other sources ---
in particular, the HCG catalogue and from Zabludoff \& Mulchaey's
(1998) sample of poor groups --- lend support to this view.  For
example, both of these other samples have been shown to have galaxy
populations skewed toward early types (Hickson 1982, Zabludoff \&
Mulchaey 1998).  More interesting, however, is the growing body of
evidence, both in the far-infrared (Allam 1998) and in H$\alpha$
(Iglesias-P\'{a}ramo \& V\'{\i}lchez 1999), that the global star
formation rates within HCGs are, on average, not enhanced relative to
field samples of similar morphological mix.  Indeed,
Iglesias-P\'{a}ramo \& V\'{\i}lchez even note a marginally significant
locus of HCG spiral galaxies of particularly low H$\alpha$ emission in
their Fig.~4; these HCG spirals may correspond to our LCRS CG sample
of low-SFR late-type galaxies.

Therefore, our initial hypothesis --- that interaction-induced
starbursts dominate the global SFR in LCRS CGs --- fails.  Although
starbursts are no doubt important, other factors prevail to yield a
net depression in the SFR in CG environments.  Much of this effect is
merely due to the high fraction of early-type galaxies in CGs, but at
least some of it is likely due to dampened activity in late-type
galaxies; this second mechanism indicates that gas stripping
mechanisms may play a role in CG environments.

\acknowledgments We thank the referee for many useful comments.  This
work was supported by the U.S. Department of Energy under contract
No.\ DE-AC02-76CH03000.  HL acknowledges support provided by NASA
through Hubble Fellowship grant \#HF-01110.01-98A awarded by the Space
Telescope Science Institute, which is operated by the Association of
Universities for Research in Astronomy, Inc., for NASA under contract
NAS 5-26555.

\clearpage

\begin{deluxetable}{cc|cc|cc|cc}
\footnotesize 
\tablewidth{0pt}
\tablecaption{The EW(\ion{O}{2}) of LCRS galaxies in different environments
\label{tab:EW}}
\tablehead{
\multicolumn{2}{c|}{Galaxies}&
\multicolumn{6}{c}{\emph{EW(\ion{O}{2})\tablenotemark{a} }} \nl
\tableline
\colhead{Environment}&\colhead{Total No.}& 
\multicolumn{2}{|c|}{\emph{NEM}} &\multicolumn{2}{c|}
{\emph{WEM}} & \multicolumn{2}{c}{\emph{SEM}} \\
& & \multicolumn{2}{c|}{\emph{EW$<$5 $\AA$}} &\multicolumn{2}{c|}
{\emph{5$\AA$ $\leq$EW$<$20 $\AA$}} & 
\multicolumn{2}{c}{\emph{EW$\geq$20 $\AA$}}
}
\startdata 
Compact Group &   104 &   72 & ($69.2\% \pm 8.2\%$) &   27 & ($26.0\% \pm 5.0\%$) &    5 & ($4.8\% \pm 2.2\%$) \nl
Loose Group   &  6612 & 4312 & ($65.2\% \pm 1.0\%$) & 1892 & ($28.6\% \pm 0.7\%$) &  408 & ($6.2\% \pm 0.3\%$) \nl
Field         & 12915 & 6804 & ($52.7\% \pm 0.6\%$) & 4905 & ($38.0\% \pm 0.5\%$) & 1206 & ($9.3\% \pm 0.3\%$) \nl
\tablenotetext{a}{The equivalent widths (EW) are classified as no
 emission (NEM; EW$<$5 $\AA$), weak emission (WEM; 5$\AA$
 $\leq$EW$<$20 $\AA$), and strong emission (SEM; EW$\geq$20 $\AA$).}
\enddata
\end{deluxetable}

\clearpage

\begin{deluxetable}{ccccc}
\tablecaption{The Concentration Index C of LCRS galaxies. \label{tab.C}} 
\tablewidth{0pt}
\tablehead{
\colhead{Galaxies Environment}&\colhead{Total No.} 
&\colhead{mean}&\colhead{ median}}
\startdata 
Compact Group & 86  & 0.324 $\pm$ 0.009 & 0.302 \nl
Loose Group   &4528 & 0.303 $\pm$ 0.001 & 0.298 \nl
Field         &8287 & 0.287 $\pm$ 0.008 & 0.28  \nl
\enddata
\end{deluxetable}

\clearpage

\clearpage

\clearpage

\clearpage

\clearpage

\end{document}